\documentclass[aps,prx,twocolumn]{revtex4-1}
\usepackage{amsmath}
\usepackage{graphicx}
\usepackage{multirow}
\usepackage{float}
\usepackage{color}
\usepackage{ulem}
\usepackage{dcolumn}
\usepackage{amssymb}

\begin{document}

\title{Strong-coupling superconductivity and weak vortex pinning in Ta-doped CsV$_{3}$Sb$_{5}$ single crystals}

\author{Jinyulin Li$^{1}$, Wei Xie$^{1}$, Jinjin Liu$^{2,3}$, Qing Li$^{1}$, Xiang Li$^{2,3,4}$, Huan Yang$^{1,*}$, Zhiwei Wang$^{2,3,4,\dag}$, Yugui Yao$^{2,3}$, and Hai-Hu Wen$^{1,\ddag}$}

\affiliation{$^1$ National Laboratory of Solid State Microstructures and Department of Physics, Collaborative Innovation Center of Advanced Microstructures, Nanjing University, Nanjing 210093, China}

\affiliation{$^2$ Key Laboratory of Advanced Optoelectronic Quantum Architecture and Measurement (MOE), School of Physics, Beijing Institute of Technology, Beijing 100081, China}

\affiliation{$^3$ Beijing Key Lab of Nanophotonics and Ultrafine Optoelectronic Systems, Beijing Institute of Technology, Beijing 100081, China}

\affiliation{$^4$ Material Science Center, Yangtze Delta Region Academy of Beijing Institute of Technology, Jiaxing, 17 314011, China}

\begin{abstract}
By measuring magnetizations of pristine and Ta-doped CsV$_{3}$Sb$_{5}$ single crystals, we have carried out systematic studies on the lower critical field, critical current density, and equilibrium magnetization of this kagome system. The lower critical field has been investigated in the two typical samples, and the temperature dependent lower critical field obtained in Ta-doped sample can be fitted by using the model with two $s$-wave superconducting gaps yielding the larger gap of $2\Delta_{s1}/k_\mathrm{B}T_\mathrm{c}=7.9\;(\pm1.8)$. This indicates a strong-coupling feature of the V-based superconductors. The measured magnetization hysteresis loops allow us to calculate the critical current density, which shows a very weak bulk vortex pinning. The magnetization hysteresis loops measured in these two kinds of samples can be well described by a recently proposed generalized phenomenological model, which leads to the determination of many fundamental parameters for these superconductors. Our systematic results and detailed analysis conclude that this V-based kagome system has features of strong-coupling superconductivity, relatively large Ginzburg-Landau parameter and weak vortex coupling.
\end{abstract}
\maketitle

\section{Introduction}
Since the discovery of superconductivity in the kagome compounds $A$V$_{3}$Sb$_{5}$ ($A$ = K, Rb, or Cs) \cite{CsV3Sb5,KV3Sb5,RbV3Sb5}, these new materials have raised extensive interests owing to their exotic physics besides superconductivity. Firstly, the charge density wave (CDW) order is observed to coexist with the superconductivity in these materials \cite{CsV3Sb5,KV3Sb5,RbV3Sb5}; however, the exact origin of the CDW state is still under debate \cite{theoryBHYan,JiangpingHu.CFP,theoryFernandes,ARPES.YZhang,ARPES.DWShen,ARPES.XJZhou}. Secondly, the topological nontrivial state is evidenced to exist in these materials \cite{CsV3Sb5,SdH.HCLei,ARPES.Sato}. A robust zero-bias conductance peak is observed in the vortex core on the Cs $2\times2$
surface of CsV$_{3}$Sb$_{5}$, and this zero-bias peak is regarded as the possible Majorana zero mode \cite{STM.XHChen.prx}. Thirdly, there are some experiments supporting the argument of symmetry breaking in these materials. Below the CDW transition temperature, an in-plane nematic electronic state is observed in CsV$_{3}$Sb$_{5}$ by electronic transport measurements \cite{YingXiang,natureXHChen}, which is consistent with the anisotropic intensity of the CDW order in CsV$_{3}$Sb$_{5}$ and its sister compound from scanning tunnelling microscope (STM) measurements \cite{STM.Hasan,STM.Zeljkovic,rotationKV3Sb5,HuazhouLi}. A natural explanation is that the three-dimensional $2\times2\times2$ or $2\times2\times4$ CDW order naturally breaks the in-plane symmetry in these materials \cite{STM.XHChen.prx,CDW.HardXRay,CDW.QuantumOscillation}. Furthermore, the symmetry in the off-plane direction is also complex. A giant anomalous Hall conductivity is observed in this family \cite{AHE1,AHE2}, and a theoretical work of the chiral flux current phase is proposed to explain this effect \cite{JiangpingHu.CFP}. Meanwhile, an unconventional chiral charge order is suggested to exist in these materials \cite{STM.Hasan,eMChA}.

Since $A$V$_{3}$Sb$_{5}$ is a compound containing the $3d$ transition element of vanadium, the superconductivity in this system is likely to be unconventional. Therefore, the superconducting pairing attracts great interests in these materials. The superconducting gap is argued to be nodal from the thermal conductivity measurement in CsV$_{3}$Sb$_{5}$ \cite{thermal.SYLi}; however, other experimental observations support the existence of the nodeless gap \cite{TDO,NQR1}. The STM measurements reveal a superconducting gap $\Delta$ of about 0.5 meV in CsV$_{3}$Sb$_{5}$ \cite{STM.HongjunGao,STM.XHChen.prx}. This gap value corresponds to the ratio $2\Delta/k_\mathrm{B}T_\mathrm{c}$ of about 5 which is larger than the 3.53 expected by the weak-coupling Bardeen-Cooper-Schrieffer (BCS) theory \cite{STM.HongjunGao}. Meanwhile, the multi-gap feature is observed in CsV$_{3}$Sb$_{5}$ based on the STM data measured at a much lower temperature, and the absence of in-gap states induced by non-magnetic impurities suggests the sign-preserving gap(s) in this material \cite{multibandSTM}. The multi-gap superconductivity is also proved by the penetration depth \cite{TDO,uSR.HCLei} and lower critical field \cite{XLDong} measurements in CsV$_{3}$Sb$_{5}$.

In order to increase the critical temperature $T_\mathrm{c}$, applying pressure and chemical doping are two effective methods. Up to now, the highest value of $T_\mathrm{c}$ can be enhanced to about 4.5 K in the Sn- \cite{Sndope} or the Nb-doped \cite{Nbdoped} CsV$_{3}$Sb$_{5}$ single crystals. In materials with coexisting superconducting and CDW phases, the superconductivity is usually enhanced when the CDW is suppressed. However, in CsV$_{3}$Sb$_{5}$, although the applied pressure or the chemical doping can suppress the CDW state, their influence to $T_\mathrm{c}$ is non-monotonic. The second superconducting dome appears near the pressure \cite{pressure.JGChen,pressure.YPQi,pressure.XLChen} or the Ti- and Sn-doping level \cite{Tidope,Sndope} where the CDW phase is completely suppressed. This fact suggests the unusual competition and the complex interplay between superconductivity and CDW in CsV$_{3}$Sb$_{5}$.

In this paper, we report comparative studies on the magnetization data measured in the single crystals of CsV$_{3}$Sb$_{5}$ and Cs(V$_{1-x}$Ta$_{x})_{3}$Sb$_{5}$ with $x\approx0.14$. We obtain several superconducting parameters of the pristine and Ta-doped samples, including the lower and upper critical fields, the Ginzburg-Landau parameter, the penetration depth, and the coherence length.

\section{Experimental method}

The pristine and Ta-doped CsV$_{3}$Sb$_{5}$ single crystals were grown by the self-flux method \cite{CsV3Sb5,PRM2019.AV3Sb,ZWWangPRB,Tadoped}. Here we choose the Cs(V$_{1-x}$Ta$_{x})_{3}$Sb$_{5}$ single crystal with the elemental doping level of $x\approx0.14$, and this is nearly the highest doping level $x_\mathrm{max}$ we can achieve. Since $T_\mathrm{c}$ increases almost monotonously with $x$ when $x\leq x_\mathrm{max}=0.14$ \cite{Tadoped}, the single crystal with $x = 0.14$ has the highest $T_\mathrm{c}$ so far when compared to other samples with a lower doping level. The magnetization measurements were carried out in a SQUID-VSM (Quantum Design) with the lowest temperature of 1.8 K. The frequency of the vibration is 13.1 Hz. The magnet was degaussed before a new round of measurement for the lower critical field ($H_\mathrm{c1}$). During the measurements of $H_\mathrm{c1}$ of undoped and Ta-doped samples and the magnetization hysteresis loops (MHLs) of CsV$_{3}$Sb$_{5}$, since the field range is rather narrow, each time of the magnetization measurement was carried out when the target magnetic field was set precisely. During the measurements of MHLs of the Ta-doped sample, the magnetization measurement was carried out while the magnetic field was sweeping. The background signal of the MHL was measured for the sample holder without any samples in the same field varying condition, and the MHLs shown in this work are the results after subtracting the background. The resistivity measurement was carried out by using the standard four-probe method in a PPMS (Quantum Design) with the lowest temperature of 1.9 K. In all measurements, the magnetic field is applied along the $c$ axis of the single crystal.

\section{Results}

\subsection{Characterization of superconducting transition}\label{s1}

\begin{figure}
\includegraphics[width=9cm]{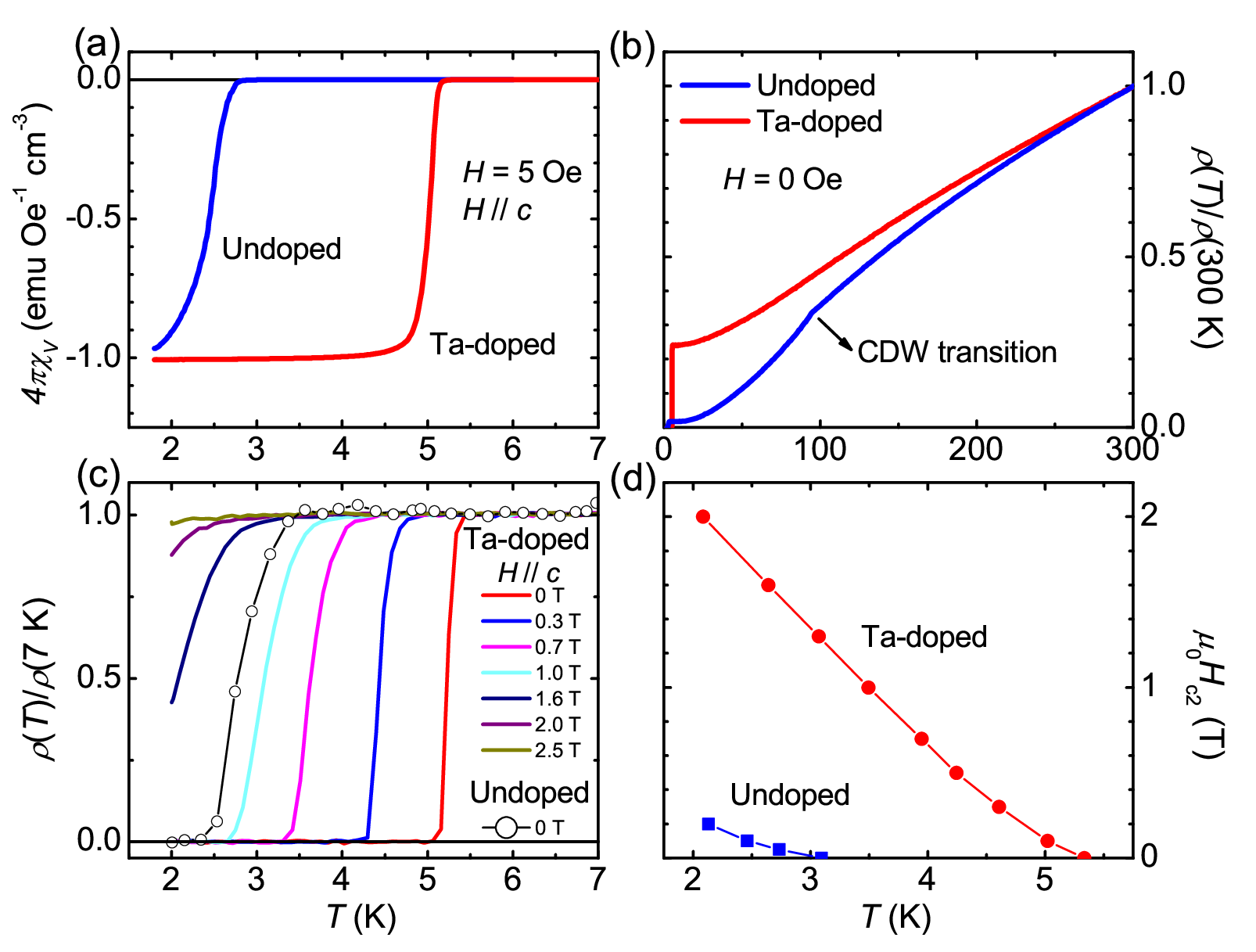}
\caption{(a) Temperature dependence of the magnetic susceptibility measured in undoped and Ta-doped single crystals, and the measurements are both carried out in the zero-field-cooled process. Here the demagnetization factor $N_\mathrm{exp}$ is considered when we calculate the magnetic susceptibility, and $N_\mathrm{exp}$ is derived from the linear fitting to the initial $M$-$H$ curve (see main text in Part~\ref{s2}). (b) Temperature dependence of the normalized resistivity measured in the pristine and the Ta-doped samples at temperatures from 2 K to 300 K. (c) Temperature dependent normalized resistivity near the superconducting transition measured in the Ta-doped sample under different magnetic fields and in CsV$_{3}$Sb$_{5}$ at 0 T. (d) Temperature dependent upper critical field of the pristine and the Ta-doped samples. The upper critical field of CsV$_{3}$Sb$_{5}$ is taken from the in-plane resistivity data in Ref.~\cite{YingXiang}.} \label{fig1}
\end{figure}

Figure~\ref{fig1}(a) shows the temperature dependent magnetic susceptibility measured in Ta-doped and undoped CsV$_{3}$Sb$_{5}$ both in the zero-field-cooled mode under the applied magnetic field of 5 Oe. The magnetic susceptibility of CsV$_{3}$Sb$_{5}$ shows obvious temperature dependent behavior below the superconducting transition temperature, which may suggest that the magnetic field of 5 Oe is enough to penetrate the sample at the edge at 1.8 K if the demagnetization factor is considered. However, the magnetic susceptibility of Ta-doped sample shows a plateau with the value about $-1$ in the low temperature region, indicating a perfect diamagnetism at 1.8 K and under the same field of 5 Oe. The difference of the $M$-$T$ curves measured in two samples suggests an obvious enhancement of the critical temperature $T_\mathrm{c}$ with the Ta doping, and the lower critical field may also be enhanced simultaneously. Figure~\ref{fig1}(b) shows the temperature dependent normalized resistivity of the two samples; both curves are normalized by the corresponding resistivity measured at 300 K. One can also see the obvious enhancement of $T_\mathrm{c}$ by the Ta doping. Meanwhile, the Ta doping leads to a stronger impurity scattering in the sample and results in a decrease of $RRR\equiv\rho(T=300\;\mathrm{K})/\rho(T=6\;\mathrm{K})$ from $59$ in the undoped sample to $4.2$ in the Ta-doped sample. The kink feature is supposed to be associated with the CDW transition in the temperature dependent resistivity curve measured in CsV$_{3}$Sb$_{5}$. However, this feature is disappeared in the resistivity curve measured in doped sample, which suggests a complete suppression of the CDW order by the Ta doping at this doping level. The evolution of the CDW state and the superconductivity by the Ta doping can be seen in Ref.~\cite{Tadoped}. Figure~\ref{fig1}(c) shows the temperature dependence of the normalized resistivity of the Ta-doped sample measured under different magnetic fields and that of the undoped sample measured at 0 T. By using the criterion of 90\% of the normal-state resistivity, the onset transition temperature $T_\mathrm{c}^\mathrm{onset}$ at 0 T is about 3.1 K for the undoped sample and 5.3 K for the Ta-doped sample. $T_\mathrm{c}^\mathrm{onset}$ of CsV$_{3}$Sb$_{5}$ is similar to values in previous reports by using the same criterion \cite{CsV3Sb5,AHE2,uSR.HCLei}, and $T_\mathrm{c}^\mathrm{onset}$ of Ta-doped CsV$_{3}$Sb$_{5}$ is higher than the highest value of about 4.5 K in the Sn- \cite{Sndope} or the Nb-doped \cite{Nbdoped} CsV$_{3}$Sb$_{5}$ single crystals. We can also determine the upper critical field ($\mu_0H_\mathrm{c2}$) by using the criterion of 90\% of the normal-state resistivity, and Fig.~\ref{fig1}(d) shows $\mu_0H_\mathrm{c2}(T)$ curve measured in Ta-doped sample. In order to compare the results, we also plot $\mu_0H_\mathrm{c2}(T)$ of CsV$_{3}$Sb$_{5}$ \cite{YingXiang} in the same figure. One can see that the Ta doping not only increases $T_\mathrm{c}$ but also raises $\mu_0H_\mathrm{c2}(T)$ significantly.

\subsection{Lower critical field}\label{s2}

\begin{figure}
\includegraphics[width=8cm]{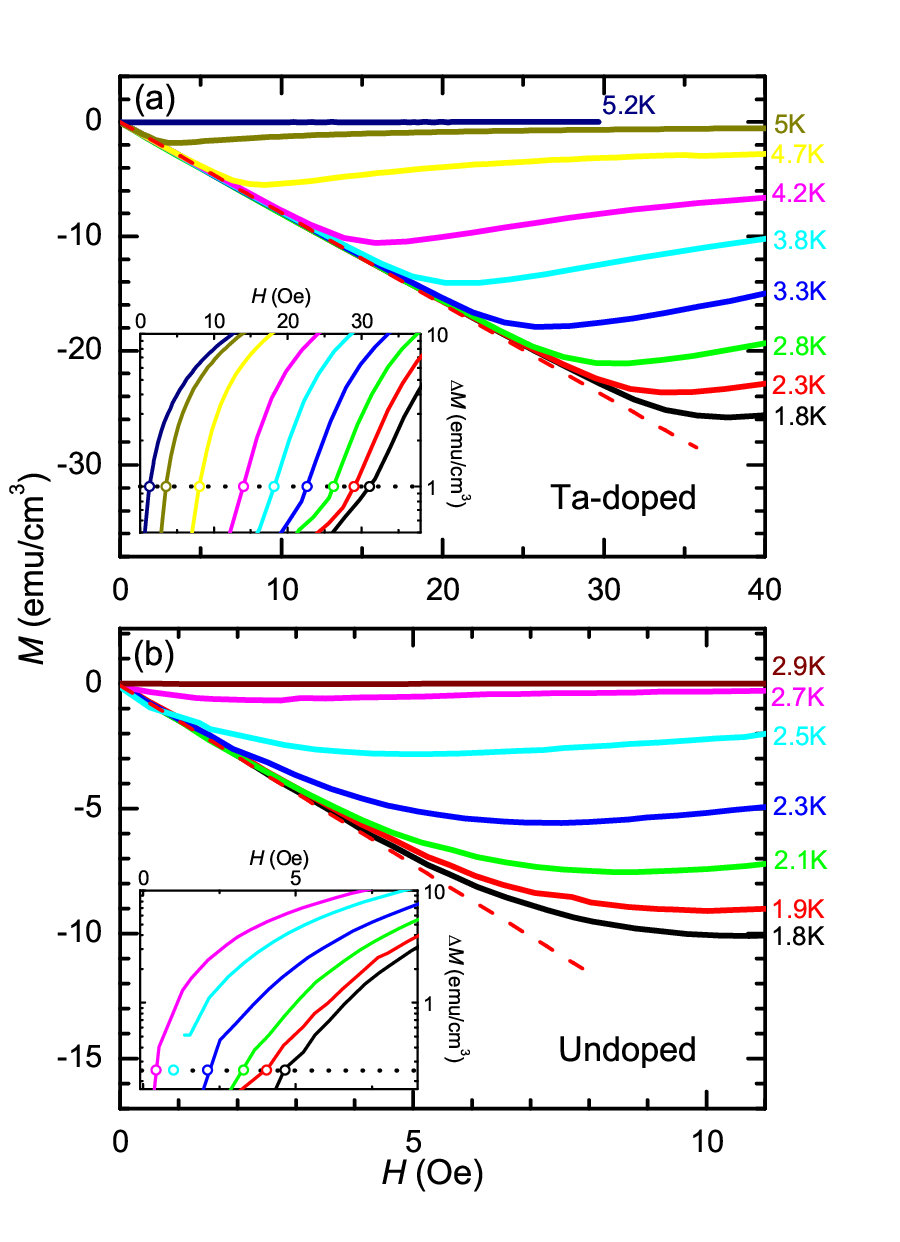}
\caption{Magnetic field dependent magnetization measured in (a) Ta-doped CsV$_{3}$Sb$_{5}$, and (b) undoped CsV$_{3}$Sb$_{5}$ at various temperatures. The dashed lines show the linear $M$-$H$ dependencies representing the perfect diamagnetic property, and the magnetization curves show the field penetration process starting from the perfect diamagnetic state. The insets show the deviation $\Delta M$ between the measured magnetization curve and the dashed line in the corresponding main panel. The dotted lines in insets are the criterions to determine $H_\mathrm{c1}$, and open circles present the determined values of $H_\mathrm{c1}$ without considering the demagnetization factor.}
\label{fig2}
\end{figure}

In the following, we focus on the magnetic field penetration process in the pristine and the Ta-doped CsV$_{3}$Sb$_{5}$ single crystals. The sample size used in the following magnetization measurement is $2.4\times2.25\times0.08$ mm$^3$ for CsV$_{3}$Sb$_{5}$, and is $1.18\times0.74\times0.058$ mm$^3$ for the Ta-doped sample. Figure~\ref{fig2} shows the field dependence of magnetization from the Meissner shielding state to the field penetration state. In this Meissner shielding state, the diamagnetic signal changes almost linearly with the magnetic field in the initial part of the $M$-$H$ curve, and then the curve deviates from the linear relationship suggesting that the field starts to penetrate into the sample in the form of vortices. It should be noted that both samples are very thin; therefore, the demagnetization factor $N$ should be considered for the analysis of the Meissner shielding effect, and then the relationship between $M$ and $H$ is $-4\pi M=H/(1-N)$. We linearly fit the initial part of $M$-$H$ curves and use the fitting results as reference lines describing the perfect diamagnetic state. The fitting lines for the two samples are shown as the dashed lines in Fig.~\ref{fig2}, and $N_\mathrm{exp}$ can be calculated from the formula $-4\pi M=H/(1-N_\mathrm{exp})$. The values of $N_\mathrm{exp}$ are 0.945 and 0.900 for undoped and Ta-doped samples, respectively. The values of demagnetization factor $N_\mathrm{geo}$ are 0.910 and 0.858 as calculated from the sizes of the two samples. However, if we use the model of an ellipsoid with three different axes, and the lengths of the three axes are set to be equal to side lengths of the sample, the calculated values of $N_\mathrm{geo}$ are 0.948 and 0.906 for undoped and Ta-doped samples, respectively. They are very close to values of $N_\mathrm{exp}$, which means that the effective shape of a very thin sample may be close to an ellipsoid. However, it should be noted that the magnetization is an output of the SQUID-VSM by fitting to the measured magnetic response of spatially drifting sample (assumed to be a magnetic dipole), which may have an error when compared to the exact value if the sample is very thin. Then, we calculate the differences between the measured $M$-$H$ curves and two dashed lines, and plot the results in the insets of Fig.~\ref{fig2}. The criterion of 1 or 0.25 emu/cm$^3$ are used to determine the deviation field in the Ta-doped or undoped samples, respectively. And these two criterions are set to be proportional to the maximum values of $-M$ in $-M(H)$ curves at 1.8 K in the two samples. Finally, the exact value of lower critical field $H_\mathrm{c1}$ can be obtained from the deviation field by multiplying the factor of $1/(1-N_\mathrm{exp})$.

\begin{figure}
\includegraphics[width=8cm]{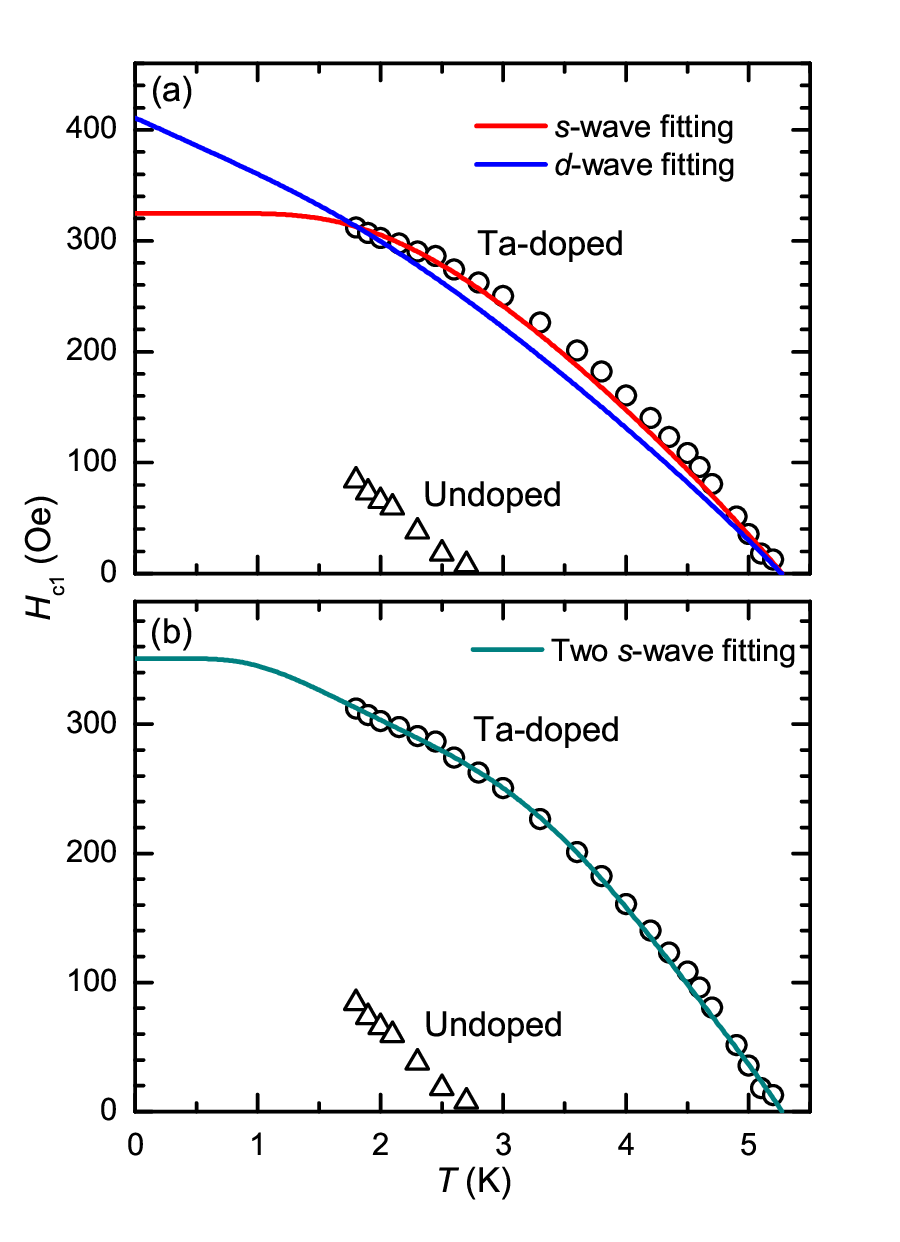}
\caption{Temperature dependent experimental data of $H_\mathrm{c1}$ obtained in the Ta-doped and the undoped CsV$_{3}$Sb$_{5}$ (symbols). The solid lines are the fitting results to $H_\mathrm{c1}(T)$ curve obtained in the Ta-doped CsV$_{3}$Sb$_{5}$. (a) Fitting results by using a single gap with the gap size of $\Delta_s=0.8$ meV, and $\Delta_d=0.98$ meV for a single $s$- or $d$-wave gap, respectively. (b) Fitting result by using the model with two $s$-wave gaps with gap sizes of $\Delta_{s1}=1.8$ meV, $\Delta_{s2}=0.43$ meV, and $x_1=0.6$ (see text). }
\label{fig3}
\end{figure}

Temperature dependence of the obtained $H_\mathrm{c1}$ are plotted in Fig.~\ref{fig3} for the Ta-doped and the undoped CsV$_{3}$Sb$_{5}$. One can see that $H_\mathrm{c1}$ is significantly enhanced by the Ta doping. The feature of $H_\mathrm{c1}(T)$ measured in CsV$_{3}$Sb$_{5}$ at temperatures down to 0.4 K has been well investigated in a previous work \cite{XLDong}. Therefore, we only analyze the data measured in Ta-doped sample. For a single-band superconductor, $H_\mathrm{c1}$ or the superfluid density $\rho_\mathrm{s}$ can be expressed as \cite{fit1,fit2},
\begin{eqnarray}
h_\mathrm{c1}(T)&\equiv&\frac{H_\mathrm{c1}(T)}{H_\mathrm{c1}(0)}=\tilde{\rho}_\mathrm{s}(T)\equiv\frac{\rho_\mathrm{s}(T)}{\rho_\mathrm{s}(0)}\nonumber\\
=1&+&2\int_0^\infty \frac{\mathrm{d}f(E,T)}{\mathrm{d}E} \frac{E}{\sqrt{E^2-\Delta(T)^2}} \mathrm{d}E.\label{eq1}
\end{eqnarray}
Here, $h_\mathrm{c1}(T)$ and $\tilde{\rho}_\mathrm{s}(T)$ are the normalized lower critical field and superfluid density ($\rho_\mathrm{s}$), $H_\mathrm{c1}(0)$ and $\rho_\mathrm{s}(0)$ are the lower critical field and the superfluid density in zero-temperature limit, $f(E,T)$ is the Fermi function, and $\Delta(T)$ is the superconducting gap function. Figure~\ref{fig3}(a) shows the best fitting results by using an $s$-wave or a $d$-wave gap. It is obvious that both fitting results can not describe the experimental data well. Since the undoped CsV$_{3}$Sb$_{5}$ is a multi-gap superconductor, we then fit the data by using the model with two $s$-wave gaps. The normalized superfluid density in a two-gap system can be expressed as $\tilde{\rho}_\mathrm{s}=x_{1}\tilde{\rho}_{\mathrm{s},s1}+(1-x_1)\tilde{\rho}_{\mathrm{s},s2}$ with $x_1$ the proportion in the total normalized $\rho_\mathrm{s}$ from the gap $\Delta_{s1}$; and the normalized lower critical field can be expressed as $h_\mathrm{c1}=x_{1}h_{\mathrm{c1},s1}+(1-x_1)h_{\mathrm{c1},s2}$. One of the best fitting results is shown in Fig.~\ref{fig3}(b), and one can see that the fitting curve coincides with the experimental data very well. Here it should be noted that the values of superconducting gaps have some uncertainties because there are several free parameters in the fitting, and the two-gap model can fit the experimental data well with fitting parameters of $\Delta_{s1}=1.8\pm0.4$ meV, $\Delta_{s2}=0.43\pm0.05$ meV. But the two-gap model is required to describe the superconducting gap in the doped sample, anyhow. This conclusion is consistent with that obtained in the undoped CsV$_{3}$Sb$_{5}$, in which two $s$-wave gaps are also required to fit the $H_\mathrm{c1}(T)$ curve \cite{XLDong}. The two-gap model is also suggested by other kinds of experiments in CsV$_{3}$Sb$_{5}$ \cite{TDO,uSR.HCLei,multibandSTM}; the obtained larger gap has the value from 0.33 to 0.57 meV, while the smaller gap has the value from 0.13 to 0.36 meV. These gap values obtained in undoped sample are much smaller than those obtained from our data in the Ta-doped sample. Since $T_\mathrm{c}$ increases from 3.1 K in the undoped sample to 5.3 K in the Ta-doped sample, the increase of the gap size is expectable.

Here the larger superconducting gap corresponds to a gap ratio of $2\Delta_{s1}/k_\mathrm{B}T_\mathrm{c}=7.9\;(\pm1.8)$, while the smaller gap corresponds to $2\Delta_{s2}/k_\mathrm{B}T_\mathrm{c}=1.9\;(\pm0.2)$. It should be noted that the expected gap ratio is about 3.53 for a single $s$-wave superconductor in the weak-coupling limit of the BCS theory, and the value of $3.53$ is just between the larger and the smaller gap ratios obtained in Ta-doped CsV$_{3}$Sb$_{5}$. Since there are multiple bands (Sb-$5p_{z}$ band and V-$3d$ branch bands) in the CsV$_{3}$Sb$_{5}$ system, and there are several saddle points near the Fermi energy, the pairing function may consist complex contributions from different scattering channels. In this case, the coexistence of a larger and a smaller gap can be explained in the framework of the BCS theory with two bands if the pairing is established between them \cite{multiband BCS}. Since it is not clear yet how strong the interband scattering is in the Ta-doped sample, we can probably discuss the situation in different ways with strong and weak interband scattering, respectively. In the case of strong interband scattering, the larger gap can be far above the BCS theoretical value of 3.53 in the weak coupling limit \cite{multiband BCS}, especially here some saddle points with high density of states may be involved in the pair-scattering. Under this particular circumstance, a smaller gap ratio is also anticipated. If the interband scattering is weak, the large gap may be formed by the pair-scattering in some individual bands, here most likely, the strong pairing is due to the intraband pair-scattering of the $d$-orbitals, and the small gap may be only passively formed. This case resembles that in MgB$_{2}$ with a strong pairing gap in $\sigma$-band and a weak one in $\pi$-band. In any case, the large gap ratio discovered here in Ta-doped CsV$_{3}$Sb$_{5}$ can be understood as a sign of strong pairing. However, we must emphasize that in a two-gap system with some impurities, like neutron irradiated MgB$_{2}$ \cite{Putti}, the interband scattering can exist and play as a fine-tuning factor to adjust the two superconducting gaps and energy-dependent density of states \cite{Ohashi}. In this regard, the momentum-resolved measurements of gap distributions are highly desired for Ta-doped CsV$_{3}$Sb$_{5}$.

\subsection{Magnetization hysteresis loop and critical current density}\label{s3}

\begin{figure}
\includegraphics[width=8cm]{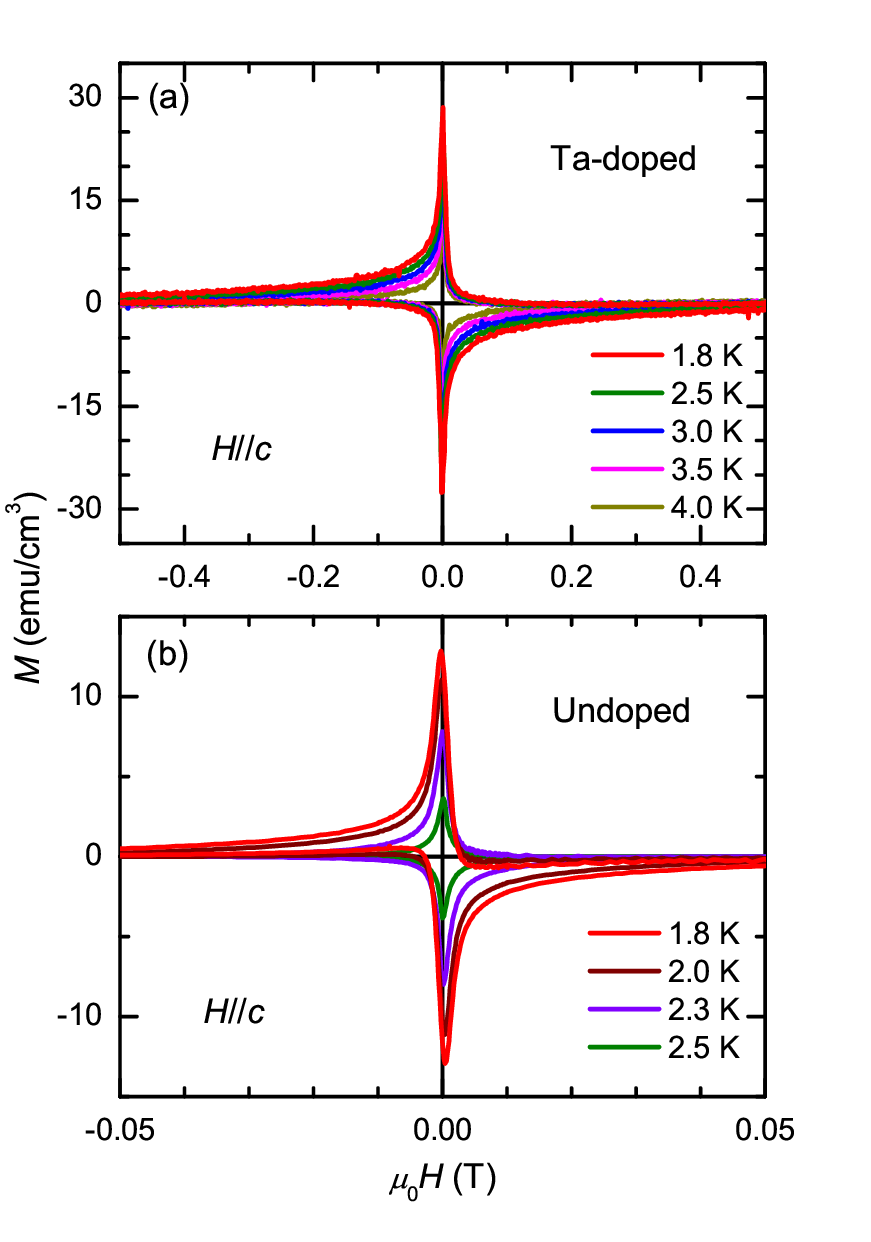}
\caption{Magnetization hysteresis loops measured in (a) Ta-doped CsV$_{3}$Sb$_{5}$ and (b) undoped CsV$_{3}$Sb$_{5}$ at different temperatures. The field ranges are different in the two figures because the irreversible fields are different for the two samples. MHLs show asymmetry with respect to the axis of $M = 0$ for both samples especially at low temperatures.}
\label{fig4}
\end{figure}

The MHL can provide fruitful information about the critical current density and the vortex pinning properties according to the Bean critical state model. Figure~\ref{fig4} shows MHLs measured in Ta-doped and undoped CsV$_{3}$Sb$_{5}$ at different temperatures. It should be noted that the irreversible field is low in CsV$_{3}$Sb$_{5}$, and especially, the field width at half the maximum is only about 3 mT for the magnetization peak near 0 T at 1.8 K. Therefore, during the measurements of MHLs in CsV$_{3}$Sb$_{5}$, the magnetization measurements are carried out when the magnetic field is fixed at the target value. In contrast, because the irreversible field is much larger in the Ta-doped sample, MHLs are measured simultaneously when the magnetic field sweeps at a rate of 20 Oe/s. Usually, MHLs are symmetric with respect to the axis of $M=0$ especially at low temperatures in superconductors with a strong bulk vortex pinning. However, one can see in Fig.~\ref{fig4} that the MHLs obtained in both samples reveal obvious asymmetry relative to the axis of $M = 0$. This fact suggests a very weak bulk pinning in both samples. In such case, the magnetization due to the bulk pinning is comparable with the equilibrium magnetization, thus the MHL is very asymmetric. One may argue that the asymmetry of the MHLs were due to the measurements with a slow field sweeping rate, thus the metastable vortex state is quickly relaxed during the measurement. In order to eliminate this concern, we also try a quick sweeping rate of 100 Oe/s for the Ta-doped sample, and the asymmetric shape of the MHL to the axis of $M=0$ is similar to that shown in Fig.~\ref{fig4}(a) although the hysteresis width does increase in high-field region. We argue that the asymmetry of MHLs is an intrinsic feature due to the relatively weak bulk pinning in this family of materials.

\begin{figure*}
\includegraphics[width=16cm]{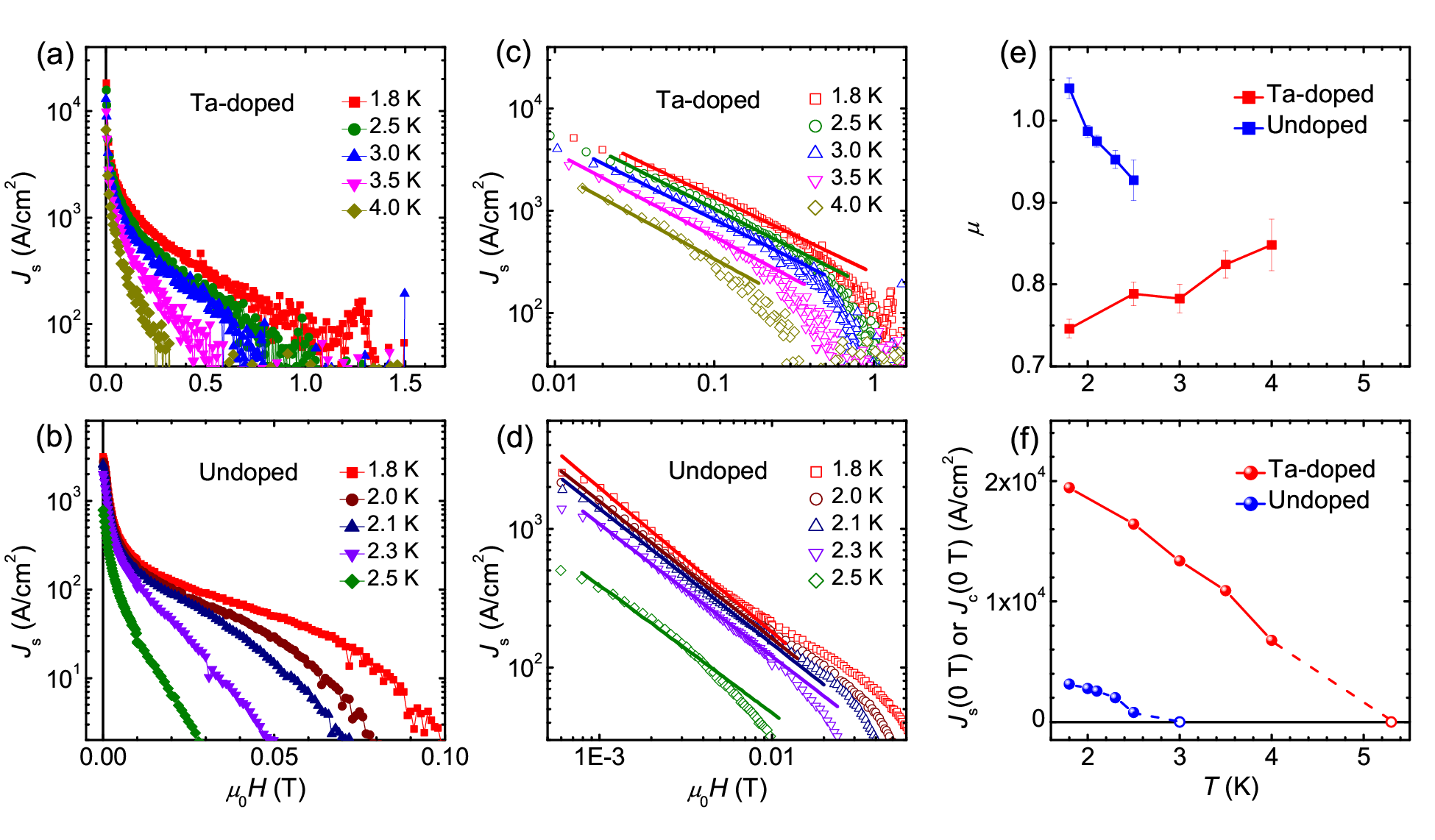}
\caption{(a,b) Semilogarithmic plots and (c,d) log-log plots of field dependent $J_\mathrm{s}$ measured in Ta-doped and undoped CsV$_{3}$Sb$_{5}$. The solid lines in (c,d) are fitting results by using the formula of $J_\mathrm{s}\propto(\mu_{0}H)^{-\mu}$. (e) Temperature dependent exponent $\mu$ derived from fittings in (c,d). (f) Temperature dependence of $J_\mathrm{s}(0\;\mathrm{T})$ obtained from (a) and (b), respectively. Empty symbols represent zero-resistance temperatures at 0 T derived from Fig.~\ref{fig1}(d), and the corresponding current density $J_\mathrm{c}$(0 T) is very close to 0 A/cm$^2$ in resistive measurements.}
\label{fig5}
\end{figure*}

It should be noted that a complete MHL consists of the field-ascending and field-descending branches. Here we define $M_\mathrm{-}$ or $M_\mathrm{+}$ as the magnetization at some magnetic fields in the field-descending or field-ascending branch of an MHL, and the hysteresis width $\Delta M$ can be defined as $\Delta M = M_\mathrm{-}-M_\mathrm{+}$. Based on the Bean critical state model \cite{Bean}, $\Delta M$ is supposed to be proportional to the critical current density $J_\mathrm{c}$. As mentioned above, the vortex state is relaxed in our measurements due to the slow field sweeping rate. Therefore, we use $\Delta M$ to calculate the transient current density $J_\mathrm{s}$ instead of the intrinsic critical current density $J_\mathrm{c}$ \cite{gis}, and $J_\mathrm{s}$ should be lower than $J_\mathrm{c}$. Then, the Bean critical state model \cite{Bean} is used to calculate $J_\mathrm{s}$, i.e., $J_\mathrm{s}=20\Delta M/[a(1-a/3b)]$. In the calculation, the unit of $\Delta M$ is emu/cm$^3$, and $a$ and $b$ are the width and length of the sample with the unit of cm. Figures~\ref{fig5}(a) and \ref{fig5}(b) show the field dependence of the calculated $J_\mathrm{s}$ in the semilogarithmic plot. As can be seen, the magnitude of $J_\mathrm{s}$ is very small in both samples, and the value decreases rapidly with the increase of the field in both samples. All these indicate a very weak bulk vortex pinning in the two samples. Having a close look, one finds that $J_\mathrm{s}$ of the Ta-doped and the undoped CsV$_{3}$Sb$_{5}$ shown in Figs. 5(a) and 5(b) decreases in different slopes with the increase of the applied magnetic field in the high-field region. However, it should be noted that the typical size of the Ta-doped single crystal is much smaller than that of the undoped sample. For instance, the volume of the CsV$_{3}$Sb$_{5}$ sample used in the magnetization measurement is one order of magnitude larger than that of the Ta-doped sample. Therefore, the discernible magnetization is much larger in the undoped sample. The decrease of $J_\mathrm{s}$ in Fig.~\ref{fig5}(a) may correspond to the larger $J_\mathrm{s}$ part ($J_\mathrm{s}>10^{2}\;$A/cm$^{2}$) in Fig.~\ref{fig5}(b). Figures~\ref{fig5}(c) and \ref{fig5}(d) show the field dependent $J_\mathrm{s}$ in the log-log plot. One can see that a power-law relationship of $J_\mathrm{s}\propto(\mu_{0}H)^{-\mu}$ is satisfied in the low-field region. This power-law decaying behavior suggests the existence of dense pinning centers with very weak pinning strength in the sample, and it can be explained by the collective pinning model as the consequence of the field dependent vortex lattice rigidity \cite{CollectivePinning,NbSe2pinning,CuBi2Se3pinning}. Figure~\ref{fig5}(e) shows the temperature dependence of the exponent $\mu$ obtained from fittings. One can see that $\mu$ is in the range of $0.93$-$1.04$ for the undoped sample, and that is $0.75$-$0.85$ for the Ta-doped sample. Since the field interval is much smaller for MHL measurements in the undoped sample than that in the Ta-doped sample, the field range is different when shown in the logarithmic coordinate. That may be the reason for different values and different temperature dependent behaviors of the exponent.

In Fig.~\ref{fig5}(f), we plot the temperature dependent $J_\mathrm{s}$ at 0 T. It should be noted that the vortex relaxation is relatively weak when the magnetic field is low \cite{WenPRL}, and then the difference of $\Delta M$ at different field sweeping rates is much smaller at 0 T than those at other finite fields (see Ref. \cite{YangPRB} for example). Therefore, $J_\mathrm{s}$ is closer to $J_\mathrm{c}$ at 0 T than at other finite fields. We also plot the zero-resistance temperatures corresponding to the $J_\mathrm{c}\approx0$ in Fig.~\ref{fig5}(f), because the current density is extremely low (about several A/$\mathrm{cm}^2$) for resistive measurements. The discontinuous evolution from $J_\mathrm{s}(T)$ from magnetization measurements to $J_\mathrm{c}$ from resistive measurement may be due to the relaxation of the vortex motion in the measurements of MHLs. At 1.8 K, $J_\mathrm{s}$(0 T) are $1.94\times10^{4}$ A/$\mathrm{cm}^2$ and $3.1\times10^{3}$ A/$\mathrm{cm}^2$ for Ta-doped and undoped CsV$_{3}$Sb$_{5}$, respectively. Although $J_\mathrm{s}$(0 T) is greatly enhanced in the Ta-doped sample, the practical value is still much lower than that of many other unconventional superconductors such as iron-based superconductors \cite{Jc.BKFA,Jc.Sm1111,Jc.CaKFeAs} and cuprates \cite{Jc.Cu1234.CPHe} whose critical current densities are normally in a range of $10^5$-$10^7$ A/cm$^2$ under the same reduced temperature of $T/T_\mathrm{c}$. The relatively low $J_\mathrm{s}$ suggests again the very weak pinning in these materials.

\subsection{MHLs fitted by a generalized phenomenological model}

\begin{figure}
\includegraphics[width=8cm]{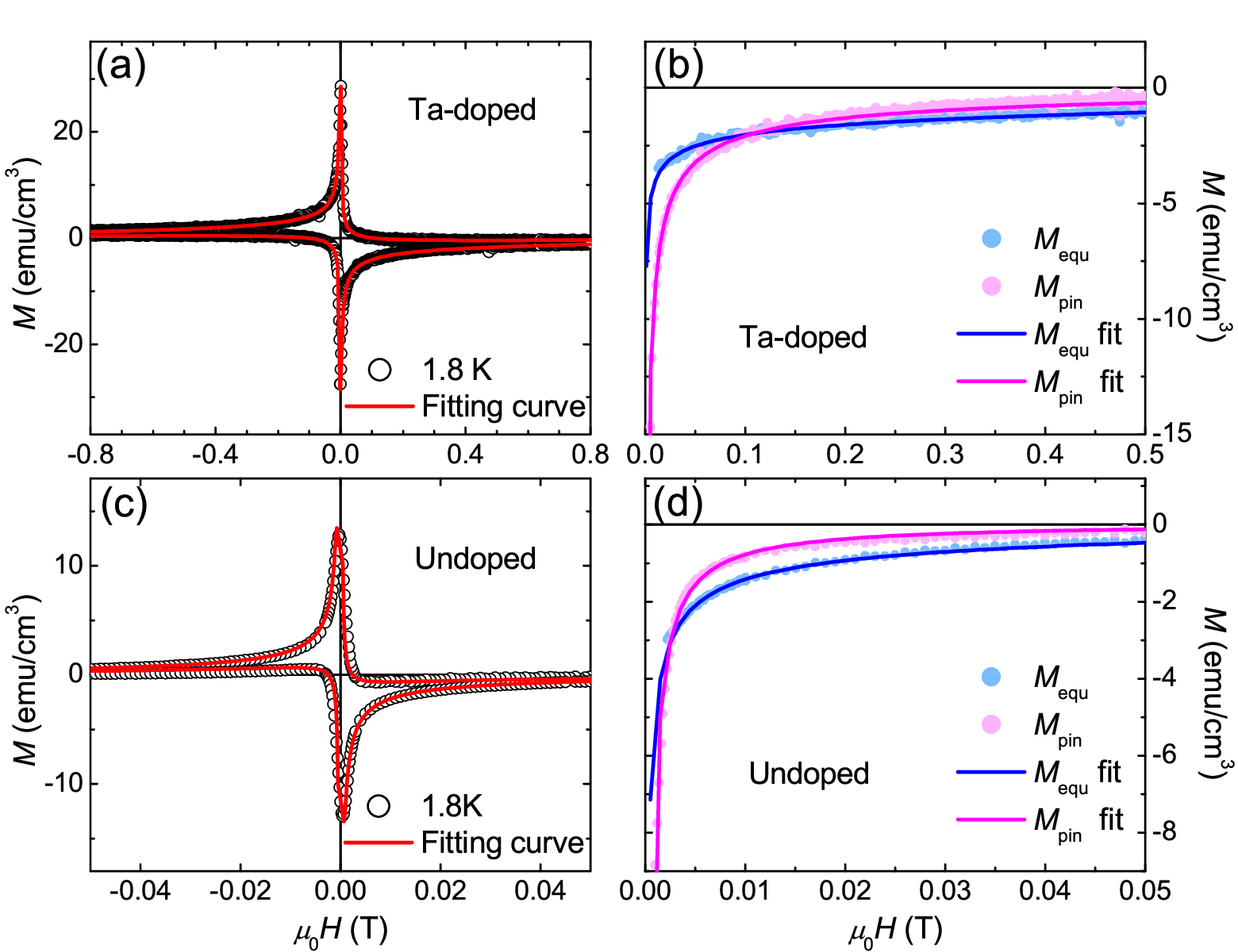}
\caption{(a),(c) Magnetization hysteresis loops (open circles) measured in (a) Ta-doped and (c) undoped CsV$_{3}$Sb$_{5}$ at 1.8 K. (b),(d) Equilibrium and non-equilibrium components of the total magnetization in (b) Ta-doped and (d) undoped CsV$_{3}$Sb$_{5}$. The solid curves in the figures show the fitting results by using the phenomenological model \cite{WeiXie}.}
\label{fig6}
\end{figure}

Recently, a generalized phenomenological model is developed to describe the magnetic field penetration and MHLs in a type-II superconductor \cite{WeiXie}, and this model can fit MHLs obtained in Nb and Ba$_{0.6}$K$_{0.4}$Fe$_2$As$_2$ very well. In the fitting process, the equilibrium and the non-equilibrium magnetization components are considered separately. Here the equilibrium part is the average of the magnetization in the field-ascending and field-descending processes of MHLs, i.e., $M_\mathrm{equ}=(M_\mathrm{+}+M_\mathrm{-})/2$; while the non-equilibrium part is the magnetization difference between the field-ascending and field-descending processes, i.e., $M_\mathrm{pin}=(M_\mathrm{+}-M_\mathrm{-})/2$. In this model \cite{WeiXie}, when the width of the sample is much larger than the penetration depth $\lambda$, $M_\mathrm{equ}$ is due to the surface screening current within the depth of $\lambda$, while the non-equilibrium magnetization $M_\mathrm{pin}$ comes from the bulk pinning with the depth from $\lambda$ to the inner. At an external magnetic field $H>H_\mathrm{c1}$, these two kinds of magnetization with the unit of $\mathrm{emu}/\mathrm{cm}^3$ can be expressed as \cite{WeiXie}

\begin{eqnarray}
4\pi (1-N)M_\mathrm{equ}=-\frac{H_\mathrm{c1}(H_\mathrm{c2}-H)}{H_\mathrm{c2}-H_\mathrm{c1}}\left(\frac{H_\mathrm{c1}}{H}\right)^{\alpha}, \label{eq2}
\end{eqnarray}

\begin{eqnarray}
M_\mathrm{pin}=J_\mathrm{c}(0\;\mathrm{T})\frac{b}{20}\frac{H_\mathrm{c2}-|H_\mathrm{e}|}{H_\mathrm{c2}}\left(\frac{H_\mathrm{c1}}
{H_\mathrm{c1}+|H_\mathrm{e}|}\right)^{\beta}. \label{eq3}
\end{eqnarray}
Here $H_\mathrm{e}=4\pi (1-N)M_\mathrm{equ}+H$ is the field at the depth of $\lambda$, and the units of all fields are Oe in equations above. In the fitting, the magnetization is modified by the demagnetization factor $N$ in order to satisfy the boundary condition of field-penetrating process. $J_\mathrm{c}(0\;\mathrm{T})$ is the zero-field critical current density and $b$ is the half-width of the sample; the corresponding units of them are A/cm$^{2}$ and cm, respectively. $\alpha$ and $\beta$ are dimensionless fitting parameters which reflect field decaying rates of the magnetization in the equilibrium and non-equilibrium processes. If other parameters are fixed, the MHL will be more asymmetric when $\alpha$ is smaller.

Figures~\ref{fig6}(b) and \ref{fig6}(d) show the calculated equilibrium and non-equilibrium components of the MHLs measured at 1.8 K. Usually the non-equilibrium component should be much larger than the equilibrium one \cite{WeiXie}; however, one can see that these two components have the similar magnitudes in Ta-doped and undoped CsV$_{3}$Sb$_{5}$. This again suggests a very weak bulk pinning in these materials. Since $H_\mathrm{c2}\gg H_\mathrm{c1}$ in Ta-doped and undoped CsV$_{3}$Sb$_{5}$, they are obviously type-II superconductors, and then we can try the newly proposed model to fit MHLs obtained in these material. The fitting results are shown in Figs.~\ref{fig6}(b) and \ref{fig6}(d) as solid curves, and one can see that they can describe the experimental data very well. With the obtained fitting results of the field dependent $M_\mathrm{equ}$ and $M_\mathrm{pin}$, we can derive MHL in the region $|H|>H_\mathrm{c1}$. The MHL in the narrow field region of $|H|<H_\mathrm{c1}$ can be calculated by the integral of field distribution in a specific penetration stage \cite{WeiXie}. The general fitting result to the MHLs are shown in Figs. 6(a) and 6(c) as solid curves, and they are consistent with the experimental data. The fitting parameters are $H_\mathrm{c1}=10$ Oe ($N$ not considered), $\mu_0H_\mathrm{c2}=2.3$ T, $J_\mathrm{c}(0\;\mathrm{T})=2\times10^{4}$ A/cm$^2$, $\alpha=0.28$, and $\beta=0.61$ for Ta-doped CsV$_{3}$Sb$_{5}$; and $H_\mathrm{c1}=5$ Oe ($N$ not considered), $\mu_0H_\mathrm{c2}=0.18$ T, $J_\mathrm{c}(0\;\mathrm{T})=3\times10^{3}$ A/cm$^2$, $\alpha=0.52$, and $\beta=1.0$ for undoped CsV$_{3}$Sb$_{5}$. The values of critical fields are similar to those obtained from the other measurements mentioned above, which verifies the validity of our model.

\subsection{Determination of superconducting parameters}

The determination of fundamental parameters such as critical fields and the Ginzburg-Landau parameter are very important for new superconductors. We obtained the values of $H_\mathrm{c1}$ and $H_\mathrm{c2}$ from the magnetic penetration and resistive measurements in Parts~\ref{s1} and \ref{s2}, thus the Ginzburg-Landau parameter $\kappa$ can be calculated from the formula of $H_\mathrm{c1}/H_\mathrm{c2} = \ln\kappa/2\kappa^2$ according to the Ginzburg-Landau theory. In addition, the penetration depth $\lambda_{ab}$ and the coherence length $\xi_{ab}$ can be deduced, i.e., $H_\mathrm{c1} = {\Phi_{0}}\ln\kappa/{4\pi\lambda^{2}}$ for $\lambda$, and $H_\mathrm{c2}=\Phi_{0}/2\pi\xi^{2}$ for $\xi$. Here $\Phi_{0}$ is the magnetic flux quantum. We can also calculate the depairing current density based on the formula of $J_\mathrm{d}=\Phi_0/3\sqrt{3}\pi\mu_0\xi\lambda^2$. Table~\ref{table1} lists the superconducting parameters obtained in this work, and all parameters are the values at the temperature of 1.8 K. Here values of $\mu_{0}H_\mathrm{c2}$ at 1.8 K are obtained from linear extrapolations of $\mu_{0}H_\mathrm{c2}(T)$ curves in Fig.~\ref{fig1}(d). And other correlated parameters are derived from these values. The value of the penetration depth determined here in CsV$_{3}$Sb$_{5}$ is consistent with that obtained from the measurement using a technique based on tunneling diode oscillator \cite{TDO}. These fundamental superconducting parameters help us to have an in-depth understanding on superconductivity in V-based superconductors. The large value of $\kappa$ confirms that the Ta-doped and undoped CsV$_{3}$Sb$_{5}$ samples are typical type-II superconductors. With Ta doping, the critical current density and critical fields increase simultaneously with an increase of $T_\mathrm{c}$. Meanwhile, the penetration depth and the coherence length decrease with the Ta doping. The possible reason may be the completely suppression of the CDW phase, and the superconductivity is enhanced.

\begin{table*}
\caption{Superconducting parameters of Ta-doped and undoped CsV$_{3}$Sb$_{5}$ at 1.8 K.}
\begin{tabular}{c|c|c|c|c|c|c|c}
\hline
 & $J_\mathrm{d}^{\parallel c}(0\;\mathrm{T})$ (A/cm$^2$) & $J_\mathrm{s}^{\parallel c}(0\;\mathrm{T})$ (A/cm$^2$) & $H_\mathrm{c1}^{\parallel c}$ (Oe) & $\mu_0H_\mathrm{c2}^{\parallel c}$ (T) & $\kappa$ & $\lambda_{ab}$ (nm) & $\xi_{ab}$ (nm) \\
\hline
Cs(V$_{0.86}$Ta$_{0.14}$)$_{3}$Sb$_{5}$ ($T_\mathrm{c}=5.3$ K) & $7.2\times10^7$ & $1.9\times10^4$ & 312 &  2.2  & 8.7  & 106.8 & 12.2 \\
\hline
CsV$_{3}$Sb$_{5}$ ($T_\mathrm{c}=3.1$ K) & $9.1\times10^6$ & $3.1\times10^3$ & 84 & 0.3 & 5.5 & 182.7 & 33.1 \\
\hline
\end{tabular}\label{table1}
\end{table*}

\section{Discussion}

Through the fittings to the $H_{c1}(T)$ data of the two systems, we find that the superconducting gap is strongly enhanced by Ta doping. From the resistivity and magnetization data, we find that the CDW transition in the doped sample is completely suppressed. This clearly indicates a competition between superconductivity and the CDW order. The ratio of $2\Delta_{s1}/k_\mathrm{B}T_\mathrm{c}=7.9\;(\pm1.8)$ for the Ta-doped sample puts the system to the class of strong-coupling superconductors.

From the magnetization measurements, we find that the critical current density is very low in Ta-doped and undoped CsV$_{3}$Sb$_{5}$, i.e., in the order of $10^3$ to $10^4$ A/cm$^2$ at 1.8 K under the self-field. There are two possible reasons for the weak pinning effect: a relatively low superfluid density, or a low density of impurities which act as the pinning centers. However, we argue that a very low superfluid density is unlikely in this family of materials. Based on the London expression for the superfluid density $n_\mathrm{s}={m^{\ast}}/{\mu_{0}\lambda^{2}e^{2}}$, we can obtain the value in use of $\lambda$, here $m^{\ast}$ is the effective electron mass, $\mu_{0}$ is the permeability of vacuum, and $e$ is the electric charge. The estimated $n_\mathrm{s}\approx1.1\times10^{21}\;\mathrm{cm}^{-3}$ for CsV$_{3}$Sb$_{5}$ if we take the effective mass $m^{\ast}\approx1.3m_0$ from theoretical calculation \cite{effectivemass}. The value of the superfluid density is smaller than the charge carrier density of $1.6\times10^{22}\;\mathrm{cm}^{-3}$ obtained from other experiments \cite{pressure.XLChen}; however, such value is not small enough to support a diluted superfluid density in this family of materials. Therefore, it seems probable that the very low critical current density is due to the very weak bulk vortex pinning. Based on the collective pinning theory \cite{collectivepinning}, the critical current density $J_\mathrm{c}$ has a relation with $J_\mathrm{d}$ as $J_\mathrm{c}\sim J_\mathrm{d}\left({f_\mathrm{pin}^2 n_\mathrm{i}}/{\epsilon_0^2}\right)^{2/3}$, where $f_\mathrm{pin}$ is the individual pinning force, $n_\mathrm{i}$ is the impurity density, and $\epsilon_0$ is the energy scale of the vortex interaction. Here values of $J_\mathrm{d}$ are about $7\times10^7$ and $9\times10^6$ A/cm$^2$ at 1.8 K for Ta-doped and undoped CsV$_{3}$Sb$_{5}$, respectively. These values are comparable to $6\times10^7$ A/cm$^2$ in bulk MgB$_{2}$ sample \cite{MgB2.review,MgB2.PenetrationDepth} or $4\times10^7$ A/cm$^2$ in Ba$_{0.6}$K$_{0.4}$Fe$_{2}$As$_{2}$ \cite{Ba122.CoherenceLength,Ba122.PenetrationDepth} at the similar temperature. As a result, the major reason for the low critical current density is due to the very small value of $\left({f_\mathrm{pin}^2 n_\mathrm{i}}/{\epsilon_0^2}\right)^{2/3}$, perhaps mainly because of the small value of impurity density $n_\mathrm{i}$. This is evidenced by the well-arranged triangular vortex lattice observed in STM measurements \cite{multibandSTM}. By comparison, such triangular vortex lattice has also been observed in 2$H$-NbSe$_2$ \cite{vortexNbSe2}, while $J_\mathrm{c}$ is only about $2\times10^3$ A/cm$^2$ obtained in the pure single crystal at 2 K \cite{JcNbSe2}.

As mentioned above in Part~\ref{s3}, in superconductors with strong bulk pinning, MHLs are usually symmetric to the axis of $M=0$. However, the MHLs obtained in the Ta-doped and undoped CsV$_{3}$Sb$_{5}$ are very asymmetric. The symmetric magnetization component to $M=0$ in MHLs or the non-equilibrium component $M_\mathrm{pin}$ is positively associated with the bulk vortex pinning strength. In our analysis, the comparable amplitudes of the non-equilibrium and equilibrium components suggest the weak bulk pinning in this family of materials. However, it should be noted that the MHLs become more symmetric at higher temperatures in both samples based on the data shown in Fig.~\ref{fig4}. This is very different from the situation in Ba$_{0.6}$K$_{0.4}$Fe$_{2}$As$_{2}$, that MHLs become more asymmetric at higher temperature \cite{WeiXie}. With the increase of temperature, the bulk pining and the surface barrier are both weakened. However, in this family of V-based superconductors, the weakening effect to the bulk pinning due to the temperature increase seems to be smaller than that to the surface barrier, and the extremely weak bulk pinning may cause a temperature insensitive behavior of the bulk pinning effect in these materials. It is expected that the bulk pinning will be enhanced if more impurities or defects are induced in the samples.

\section{Conclusion}

In conclusion, we have carried out extensive studies on magnetizations of the high-quality single crystals of Cs(V$_{0.86}$Ta$_{0.14})_{3}$Sb$_{5}$ and CsV$_{3}$Sb$_{5}$. Being similar to CsV$_{3}$Sb$_{5}$, the Ta-doped sample is also a multiband superconductor with the larger gap ratio of $2\Delta_{s1}/k_\mathrm{B}T_\mathrm{c}=7.9\;(\pm1.8)$ and the smaller gap ratio of $2\Delta_{s2}/k_\mathrm{B}T_\mathrm{c}=1.9\;(\pm0.2)$ according to the fitting to the temperature dependent lower critical field. This indicates a strong coupling feature of the superconductivity. Some important superconducting parameters are obtained for this family of materials. It is found that the critical current density is extremely low in these materials because of the very weak bulk vortex pinning. Our results provide fruitful information of superconductivity and vortex pinning in the Ta-doped and undoped CsV$_{3}$Sb$_{5}$.

\begin{acknowledgments}
This work was supported by National Natural Science Foundation of China (Grants No. 12061131001, No. 11927809, No. 92065109, No. 11734003,  No. 11904294 and No. 11904020), National Key R\&D Program of China (Grants No. 2022YFA1403201 and 2020YFA0308800), Strategic Priority Research Program (B) of Chinese Academy of Sciences (Grant No. XDB25000000), Beijing Natural Science Foundation (Grants No. Z210006 and Z190006). Z.W. thanks the Analysis \& Testing Center at BIT for assistance in facility support.
\end{acknowledgments}

$^{*}$ huanyang@nju.edu.cn

$^{\dag}$ zhiweiwang@bit.edu.cn

$^{\ddag}$ hhwen@nju.edu.cn

\end{document}